\documentclass{kluwer}    
\usepackage[dvips]{graphicx}
\newdisplay{guess}{Conjecture}
\begin{document}                                                                                   
\begin{article}
\begin{opening}         
\title{THE X10 FLARE ON 2003 OCTOBER 29: TRIGGERED BY
MAGNETIC RECONNECTION BETWEEN COUNTER-HELICAL FLUXES?}  

\author{YU \surname{LIU}}
\institute{Institute for Astronomy, University of Hawaii, 2680 Woodlawn Drive, Honolulu, HI 96822}

\author{HIROKI \surname{KUROKAWA}}    
\institute{Kwasan and Hida Observatories, Kyoto University, Yamashina-ku,
Kyoto 607-8471, Japan }

\author{CHANG \surname{LIU}}
\institute{Center for Solar-Terrestrial Research, New Jersey Institute of
Technology, University Heights, Newark, NJ 07102-1982}

\author{DAVID H. \surname{BROOKS}, JINGPING \surname{DUN}, TAKAKO T. \surname{ISHII} }    
\institute{Kwasan and Hida Observatories, Kyoto University, Yamashina-ku,
Kyoto 607-8471, Japan }  

\author{HONGQI \surname{ZHANG}}
\institute{National Astronomical Observatories, Chinese Academy of Sciences,
Beijing 100012, China} 

\runningauthor{LIU et al.}
\runningtitle{Reconnection of Counter-helical fluxes for a major flare}

\date{October 26, 2006}

\begin{abstract}
Vector magnetograms taken at Huairou Solar Observing Station (HSOS) and Mees Solar Observatory (MSO) reveal that the super active region (AR) NOAA 10486 was a complex region containing current helicity flux of opposite signs. The main positive sunspots were dominated by negative helicity fields, while positive helicity patches persisted both inside and around the main positive sunspots. Based on a comparison of two days of deduced current helicity density, pronounced changes were noticed which were associated with the occurrence of an X10 flare that
peaked at 20:49 UT, 2003 October 29. The average current helicity density (negative) of the main sunspots decreased significantly by about 50\%. Accordingly, the helicity densities of counter-helical patches (positive) were also found to decay by the same proportion or more. In addition, two hard X-ray (HXR) `footpoints' were observed by the Reuven Ramaty
High Energy Solar Spectroscopic Imager ({\sl RHESSI}) during the flare in the
50--100 keV energy range. The cores of these two HXR footpoints were adjacent to the positions of two patches with positive current helicity which disappeared after the flare. This strongly suggested that the X10 flare on 2003 Oct. 29 resulted from reconnection between magnetic flux tubes having opposite current helicity. Finally, the global decrease of current helicity in AR 10486 by $\sim $ 50\% can be understood as the helicity launched away by the halo coronal mass ejection (CME) associated with the X10 flare.

\end{abstract}    
\end{opening}           

\section{Introduction}  
The theory of magnetic reconnection is an important tool for explaining the
trigger and energy release processes for solar flares and other eruptions.
Highly twisted structures of an emerging magnetic system are usually thought to
be an omen of high level flare productivity \cite{Ishii00, Kurokawa02}, and
the properties of the twist can be revealed partly from observations of the
vector magnetic field \cite{Leka03a, Leka03b}. Recently, helicity related
research has become a hot topic in solar physics, not only because of
questions of its origin, transfer and conservation from inside the Sun out into
the corona, but also because of its role in solar flares. Theories and
numerical simulations have thoroughly investigated and demonstrated the process
of reconnection between magnetic flux loops of different chirality
(handedness) \cite{Sakai92, Linton01}. They suggest that the collision
between two counter-helical ropes, i.e., a right- and a left-handed flux tube, is most ideally
suited for reconnection \cite{Linton01}. \inlinecite{Kusano04}, based on 3D numerical simulations, show that the reconnection between large-scale oppositely sheared loops in a coronal arcade can trigger solar flares. Observationally, the appearance of opposite sign for helicity flux in flare-productive active regions has been
noted \cite {Deng01, Liu02}, but their relationship to the reconstruction of
the magnetic field is still not clear. Thus far, there has been no reliable
evidence from observations supporting a strong tendency for
counter-helical reconnection to result in major flares.

On 2003 Oct. 29, in the decaying period of Solar Cycle 23, a large flare
classed 2B/X10 occurred in super AR 10486 which produced a series of major
flares during its passage on the solar disc. The flare started at 20:37 UT and reached its maximum at 20:49 UT seen from GOES soft X-ray flux profile. Our observations reveal that the
major flare on Oct. 29 had a close relationship with the reciprocal action
between counter-helical fluxes. It may be the first observational evidence for
counter-helical reconnection playing an important role in the triggering of a
major flare. In \S2 we detail the observations and present the results, then,
in \S3 we present a summary and discussion of the findings.

\section{Data and Observations}
The vector magnetograms were taken by the HSOS Solar Magnetic Field Telescope
(SMFT) at FeI $\lambda$
5324.19 \AA\ \cite{Deng97, Zhang03} and MSO Imaging Vector Magnetograph (IVM) at FeI $\lambda$ 6302.5 \AA\ \cite{Mickey96, LaBonte99}. The SMFT/HSOS was designed
to measure the line-of-sight component in the band wing -0.075 \AA\ off
the center and the transverse component in the line center, in
order to reduce the cross-talk effect. A new 6701-type CCD system (640 $\times$
480 pixels) was installed at HSOS in 2001 October with a field of view
(FOV) of $3.75' \times 2.81'$ resulting in a practical image scale of about
$2''$ pixel$^{-1}$. The IVM/MSO is a Stokes profile analyzing magnetograph and can supply real-time polarization data for vector magnetic field measurement. \inlinecite{Zhang03} compared in detail the photospheric transverse fields between IVM and SMFT for a
simple sunspot, and found only a small statistical difference as 3.0 degree. In this work there are 12 sets of vector magnetograms from HSOS are used, which were chosen because of relatively better seeing conditions, including 8 sets on Oct. 29 (01-07 UT) and 4 sets on Oct. 30 (01-05 UT). Moreover, 5 additional vector magnetograms were supplied by IVM/MSO observations
which covered the 2B/X10 flare that occurred from 20:37 UT to 21:01 UT on 2003 Oct. 29 when AR 10486 was located (S$15^{\circ}$, W$02^{\circ}$) on the central meridian of the solar disc. For both HSOS and Mees data the longitudinal magnetic component (${B_z}$) and the transverse components (${B_x},{B_y}$) are derived directly from Stokes parameters. The 180 degree ambiguity is resolved by using the reference potential-field method, i.e., $\overrightarrow{{B_t}}\cdot \overrightarrow{{B_t}}{_{,pot}}\geq 0$, where $\overrightarrow{{B_t}}$ is the observed transverse field vector and $\overrightarrow{{B_t}}{_{,pot}}$ is the potential-model calculated vector on the photosphere. In the data pre-processing, we set the value thresholds on the longitudinal magnetic field component ($|{B_z}|>50$ G) and on the transverse magnetic field components ($|{B_t}|>350$ G).

The 50--100 keV hard X-ray (HXR) data used in the work are provided by {\sl RHESSI} \cite {Lin02}. HXR image is integrated from
20:45:30 to 20:46:30 UT and constructed with the CLEAN alogrithm with a $9.8''$ FWHM resolution (using {\sl RHESSI} grids 3-9). The non-thermal sources
($\geq$ 20 keV) indicated by the HXR data suggest the locations of the
footpoints of the reconnected loops along which the accelerated electrons/ions
passed before colliding with the ambient solar atmosphere \cite{Lin02}.

The other data sets, including H$\alpha$ and EUV observations, are presented in \cite{Liu06}. The multi-wavelengths data in movie form are available in the electronic version of the paper. They are useful for clear demonstration of the flare core evolution and the global effects of the X10 flare event. The H$\alpha$ movie shows remote brightenings more than $2\times10^5$ km away from the main flare in eastern and southern directions. The remote brightenings were found cotemporal with the two phases of the flare emissions (before and after $\sim$ 20:49 UT), respectively, as observed at HXR and radio wavelengths. Also, coronal dimmings at EUV wavelength were found at the locus of the remote brightenings right after the flare.

Figure 1 shows an example HSOS vector magnetogram taken at 05:43 UT,
2003 Oct. 29. Note that the actual size of the active region is larger than the
observed FOV, but Figure 1 displays the main portion of the sunspot group and
contains the most important spots of interest. High shear is clearly seen in
the transverse fields at the top left of the magnetic neutral line. Another
feature of the AR is the intrusion of a slender negative flux region into the
positive field, as shown in the top region in the image, which could be one
important cause of the highly sheared morphology formed there.

As a comparison, the longitudinal current ($J_z$) distributions on Oct. 29
and Oct. 30 deduced from the vector magnetograms ($J_z={1\over \mu_0}({{\partial
B_y}\over {\partial x}}-{{\partial B_x}\over {\partial y}}$)) are shown in the
left column of Figure 2. The grey images superposed are the corresponding
longitudinal magnetograms. It seems that the current system of this region
is running out from the region surrounding the main sunspots and then downward through them. The corresponding current helicity density results ($h_c$) are shown
in the right column of Figure 2. The current helicity density is obtained from
$h_c = \mu_0 (B_z \cdot J_z)$. Obviously, the flux in the FOV is a mixture
of both positive and negative $h_c$. The main sunspots are dominated by
negative helicity density, i.e., left-handed twisted. However, some positive
$h_c$ patches, i.e., right-hand twisted, exist around or even inside the main
sunspots. These opposite-sign $h_c$ patches are long-lived features as it is shown
from Figure 3. From Oct. 29 to 30, a decrease is evident for both the
positive and negative $h_c$ (compare c-d in Figure 2). Some weaker positive
$h_c$ patches disappear in the density map of Oct. 30.

We show in Figure 3 more details of the evolution of  $h_c$ in the two days
with 12 vector magnetograms by HSOS and Mees. In Figure 3a, four long-lived positive $h_c$ patches marked with `A-D' are identified.
The region of dominant negative $h_c$ is marked with `sunspot' since it is
the umbra of the main sunspots in the white light image in Figure 1. Between Oct.
29 and 30, `A', `B' and `C' underwent obvious evolution, while on the
contrary `D' did not. Unlike `A' and `B', region `C' resides
inside one of the sunspots. As `B' disappears, `A' still exists in the $h_c$
maps of Oct. 30 even although it has greatly decreased, while the patch `C'
has changed into a weak signal and is sometimes hardly detected (e.g. Figure 3l). Figure 3i-j are the images just taken before and after the large flare by
IVM and the features marked in Figure 3a can also be identified here. Taken into
account the time difference (11 hours between Figure 3h-HSOS and Figure 3i-MSO),
some new features shown in Figure 3i could be attributed to the long time
evolution of the active region. However, the reality of a new positive $h_c$ patch arrowed in
Figure 3j is thought questionable.

The quantitative results of the $\bar h_c$ evolution are shown in Figure 4 for
the typical patches `A', `sunspot' and the positive and negative regions of the FOV,
respectively. The peak time of the X10 flare is indicated by the impulsive rise of the SXR flux
in each panel. Before the major flare, a continuous increase in $\bar h_c$
is a common tendency among the four regions. However, after the flare time,
the values of $\bar h_c$ are suddenly reduced to a relatively lower level. The scales
of these large decreases in $\bar h_c$ are 51\% (Figure 4a-c) and
44\% (Figure 4d). If the current helicity measurements based on the SFMT and IVM data are trustworthy, then such an abnormal decrease of $\bar h_c$ in the active region
should suggest that the magnetic flux loops have been untwisted in an
effective way which may have a close relationship with the X10 flare.

Figure 5a is an enlargement of Figure 3h, in which the contours of the patches
`A-C' are indicated. The {\sl RHESSI} 50-100 keV HXR contours are
superposed in white on this  $h_c$ map in Figure 5b. The HXR data are integrated over a time period of 1 minute (20:45:30-- 20:46:30
UT), and the overlying contour levels are 10\%, 20\%, 40\%, 70\% and 90\% of
the maximum. Two obvious HXR cores are adjacent to `B' and `C',
respectively. Furthermore, it is interesting to find that the
weaker footpoint corresponds to the stronger patch `C', while the stronger
footpoint corresponds to the weaker patch `B'. From the HXR movie (see electronic version), the
footpoints over `B' and `C' show a primary separating motion during the
flare and the positions of the HXR cores are a snapshot near the flare peak time. In Figure 5b, there is also one
weak HXR emission over the large positive helicity patch `A' at 20:45 UT, but
this weak emission lasted only one minute during the whole flare period
(20:37-21:01 UT), thus we think it might be a thermal effect caused by the remote
flaring ribbons.

\section{Summary and Discussion}

During the decaying period of Solar Cycle 23, in 2003 October, several X-class flares occurred in AR 10486. One of the flares, 2B/X10 on Oct. 29,
is found to be associated with significant current helicity ($h_c$) changes in the whole
active region based on the vector magnetograms taken by Huairou Solar Observing Station and Mees Solar Observatory. Three long-lived helical flux patches
($h_c > 0$) have a significantly reduced $h_c$ after the major flare. They were located either compactly around
the main sunspots or resided just inside them. In addition, two {\sl
RHESSI} HXR footpoints appeared during the flare and they had a
one to one correspondence with the two weaker helical patches (B and C). After the
flare, these two patches disappeared, and the third patch (A) was found to decay in
$\bar h_c$ by $\sim50$\%. Furthermore, the $\bar h_c$ of the main sunspots
also decreased by half. 

In fact, before the major flare, the $\bar h_c$ of the active region kept
increasing at a high rate (Figure 4), which may indicate fresh twisted flux emerging from below. As an evidence, \inlinecite{Liu06} reported flux emergence and continuous sunspot
rotation until the time of the flare in a way to increase the overall
twist of the active region field before the major flare. After the flare, the amount of
both the positive and negative helicity was found to decrease by about the
same level, $\sim$50\%, suggesting some `cancellation' between them.
\inlinecite{Liu02} reported observations of a $\delta$ spot evolution in
another super AR 9077. They found that helicity reversal of one spot in
the $\delta$ group had a close relation with their group disintegration, and that the
helicity reversal might result from the emergence of opposite-sign helical
flux. Different from the case analyzed in this paper, they could not find obvious change in $\bar h_c$ in the whole active region. In this case of the 29-30 October, both the positive and negative current helicities partly cancel each other resulting in a reduction of the twist of the whole system.

Several authors investigated in detail the relationship between vertical
current distributions and flare kernels \cite{Lin87, Canfield93}. By using
high resolution HXR data and vector magnetograms, \inlinecite {Li97} further
pointed out that non-thermal electron accelerations usually take place adjacent
to pre-existing current system observed at the photospheric level, i.e.,
flares occur between different set of magnetic connectivities. In this work, we study
the current helicity density instead of the current density for the convenience
of identification of different helical flux systems. In AR 10486, the
locations of the HXR footpoints are found to be adjacent to the helicity
patches (`B', `C', in Figure 5), confirming the previous study
\cite{Canfield93, Li97} and suggesting the reconnection occurred between
left-handed and right-handed magnetic fluxes.

Cancellation of counter-helical fluxes has been
simulated by \inlinecite{Linton01} who demonstrated how counter-helical fluxes
reconnect and release the non-potential energy. One important finding in the
simulations is that the transverse (azimuthal) fields will reconnect and then
tie the flux loops together, thus pulling them closer to each other and
allowing further reconnection to take place. This is a special feature of
reconnection of counter-helical fluxes and such a type of reconnection is the
`most efficient reconnection' for major flare production \cite{Linton01}.

One result of counter-helical reconnection is the untwisting of the flux
loops, i.e., the decrease of the transverse components, which helps
to decay the system $\bar h_c$. The sudden disappearance of the penumbra
for sunspots in AR10486 (corresponding to `B', Figure 3) was found to be
associated with the occurrence of the X10 flare \cite{Wang04, Liu05}, directly
supporting the suggestion of decay of the transverse fields. Also, the Moreton waves and the type II radio bursts in this event were generated in a wide directional range \cite{Liu06}, as one should expect if the reconnection occurred in a way to unwind the magnetic helicity. Before adding the IVM/Mees data, it could be suspicious that the substantial current helicity changes, observed not only in small spatial scale but in large scale in the active region, were possibly due to the system noises (e.g., by instrument resettings and changing seeing conditions at HSOS during a period of two days). The complementary IVM data just cover the important period before and after the large flare, and the raw data have been corrected for the terrestrial atmosphere and the instrument systematic effects. The HSOS and Mees data jointly allow us to examine the successive evolution of the active region. Therefore, based on the above discussion, it is strongly suggested that the
X10 flare event is triggered by reconnection between the counter-helical
flux systems in the corona. It may be the first observational evidence for reconnection
between counter-helical tubes triggering major flares, although
there is a large data gap between the magnetograms and the HXR data.

There was a halo CME associated with the X10 flare. The large-scale activities have been studied by \inlinecite{Liu06}. The results from the multi-wavelength full-disk data analysis indicate salient magnetic field re-organizations in the corona due to the large-scale magnetic reconnection in the corona (as evidenced by the remote flare brightenings) that gave rise to the halo CME. The active region 10486 is just the source region that launched the fast halo CME. The radio, H$\alpha$, EUV, HXR movies are available as the electronic material of the paper. 
For the global current helicity decrease in the whole active region, we think the released halo CME should have taken away significant free helicity/energy from its source region, while the magnetic cancellations between the counter-helical fluxes played a trigger mechanism for the CME initiation.

On the
other hand, as it has been shown, the vector magnetogram data in combination with
other wavelength observations is a powerful tool for solar physics research. It
is a pity that there is no vector data for the period of another flare, or
poor seeing conditions at that time, since a series of X-class flares occurred
in AR 10486. It should be noted that Linton's simulation applies in the convective zone with confined flux tubes in high plasma beta, it is unclear that whether the same physics is applicable in the solar corona. The case presented here suggests the magnetic reconnection between counter-helical flux could work in the corona. We are expecting that the Hinode will be able to supply more convincing evidence by providing continuous high spatial and temporal resolution vector magnetograms, and overcome the significant limitations of the ground-based observations.

\acknowledgements
The referee of the paper is acknowledged for his/her many helpful suggestions 
that have improved the study. We would like to thank the Huairou Solar Observing
Station and the Mees Solar Observatory for the vector magnetic field data provided. We are also grateful to the {\sl RHESSI} team for the HXR data. Specially, we are grateful to J. Li, K. D. Leka, D. Mickey and B. LaBonte for the IVM data reduction. CL acknowledges support from NSF/SHINE grant ATM 05-48952.

\begin{figure}
\centerline{\includegraphics[scale=0.7]{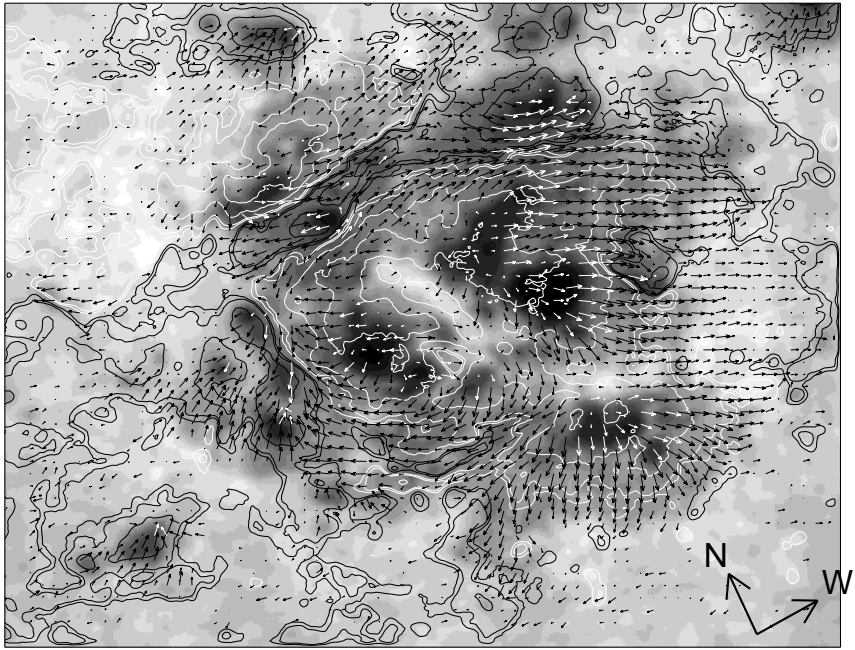}}
\caption{ HSOS vector magnetogram obtained at 05:43 UT, 2003 Oct. 29 overlying
the corresponding white light image. The white/black contours represent
positive/negative longitudinal magnetic fields, respectively. The arrows are
the transverse fields and their length is proportional to the field strength. The
magnetic contour levels are $\pm$ 150, 300, 1200, 2400 G. The FOV
is $3.5'\times2.7'$. Solar north and west are the directions indicated, and are the same in the following figures.}   
\label{fig1}  
\end{figure}

\begin{figure}
\centerline{\includegraphics[scale=0.8]{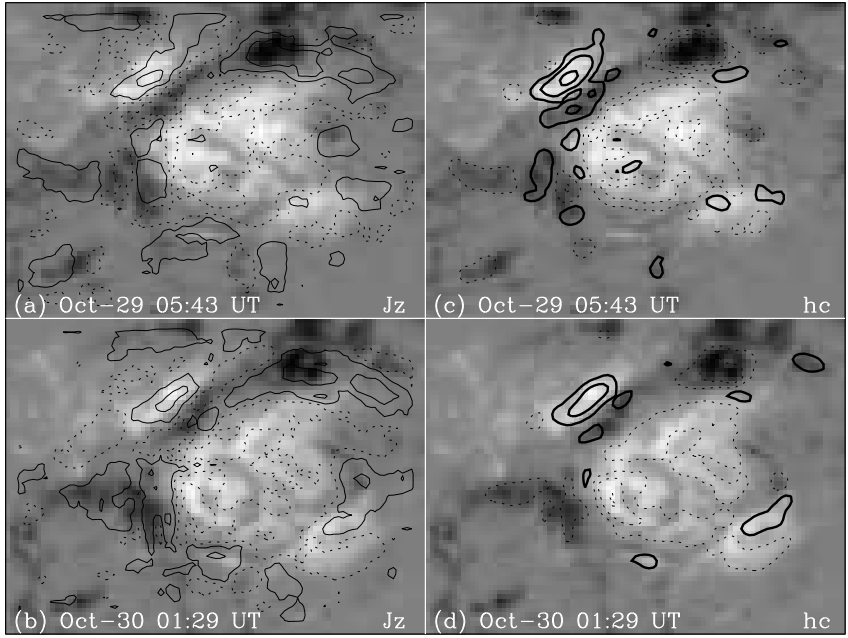}}
\caption{Comparison between Oct. 29 and 30 to examine the changes in the
longitudinal current density ($J_z$, in contours, left column) and the current
helicity density ($h_c$, in contours, right column). The grey-scale images are the
corresponding longitudinal magnetograms with positive polarity flux in {\it white}
and negative in {\it black}. In ({\bf a}) and ({\bf b}), the {\it solid black contours}
represent the upward component of the current density ($J_z >$ 0), while the
{\it dashed black contours} represent the downward components ($J_z <$ 0), their
levels are $\pm 2, 8, 20, 40 \times10^{-3} A\cdot m^{-2}$. In ({\bf c}) and
({\bf d}), the {\it thick solid black contours} represent the upward component of
current helicity density ($h_c >$ 0), while the {\it dashed black contours} represent
the downward components ($h_c <$ 0), their levels are $\pm 2, 8, 20, 40
\times10^{-2} G^{2}\cdot m^{-1}$.
\label{fig2}}
\end{figure}

\begin{figure}
\centerline{\includegraphics[scale=0.75]{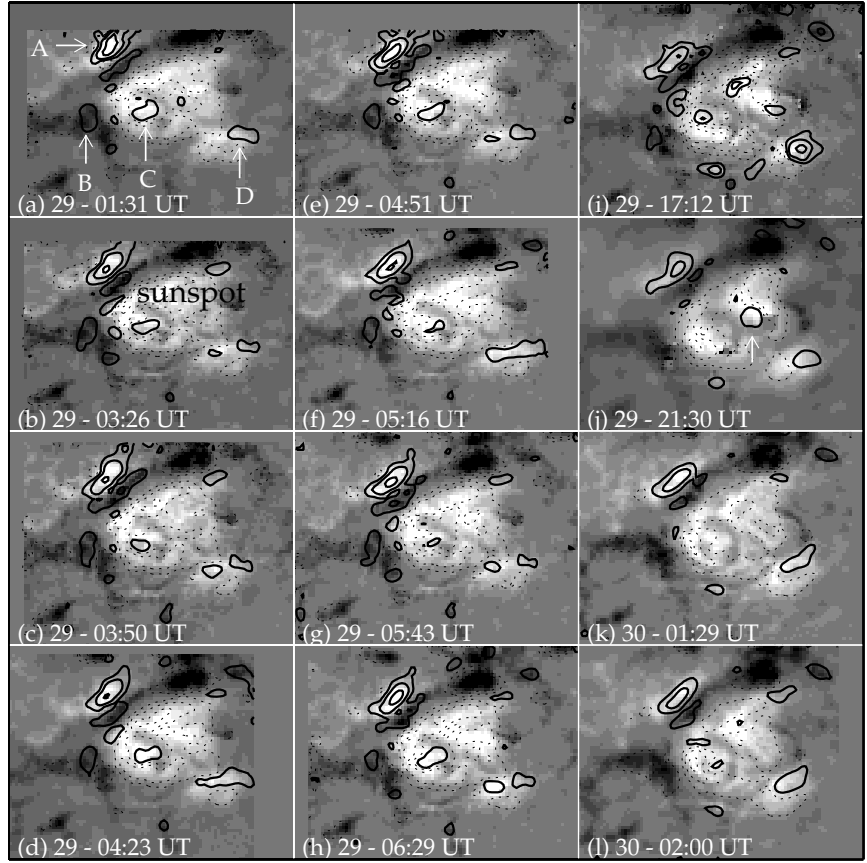}}
\caption{Evolution of the current helicity density ($h_c$) during two
days of observations. The grey-scale images superposed are the longitudinal magnetic
fields. In ({\bf a}), `A-D' are four positive helicity features pointed out by
the arrows. In ({\bf b}), `sunspot' means the area, in white,
that is dominated by the negative helical fields in which the counter-helical
feature `C' exists. Note that ({\bf i}) and ({\bf j}) were
taken by MSO and the others were by HSOS.
\label{fig3}}
\end{figure}

\begin{figure}
\centerline{\includegraphics[scale=0.7]{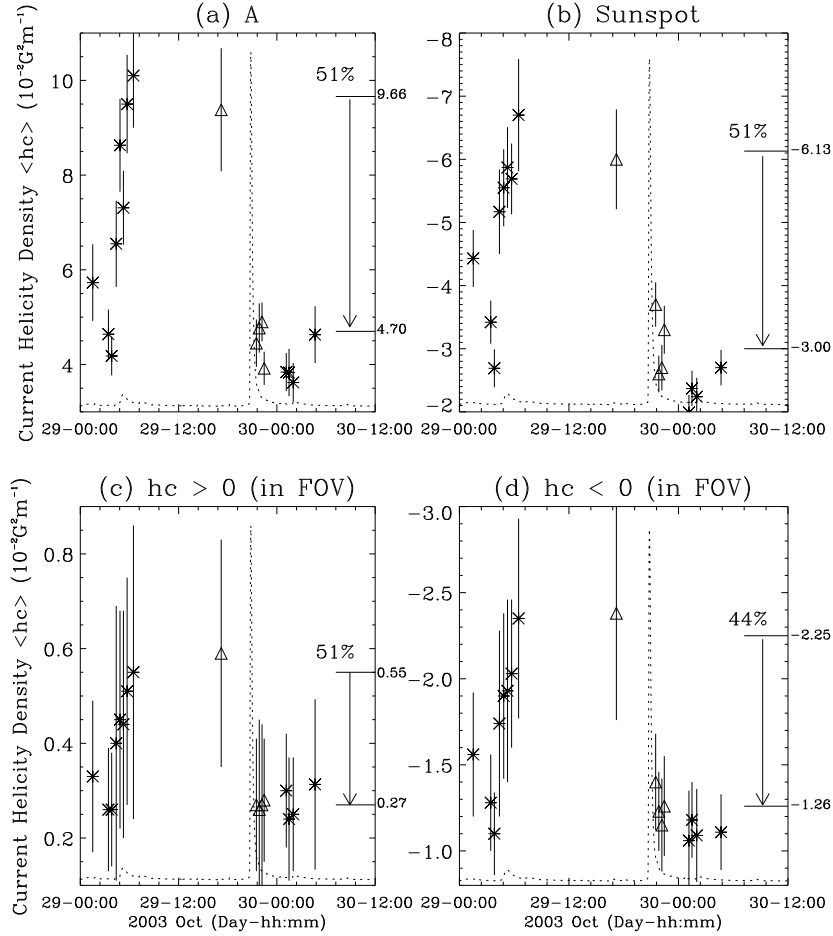}}
\caption{Measurement of average $h_c$ based on 17 magnetograms for four
different objects. Each {\it asterisk} (Huairou data) or {\it triangle} (Mees data) 
represents one magnetogram. The dotted line is {\sl GOES} SXR 1-8 \AA\ flux profile (normalized) as a function of time. The X10 flare peaked at 20:49 UT on
Oct. 29. In each panel, a rough percent estimation is shown for the sudden $\bar
h_c$ decrease between two values which are from averaging three
successive points before and after the X10 flare, respectively.
\label{fig4}}
\end{figure}

\begin{figure}
\centerline{\includegraphics[scale=0.7]{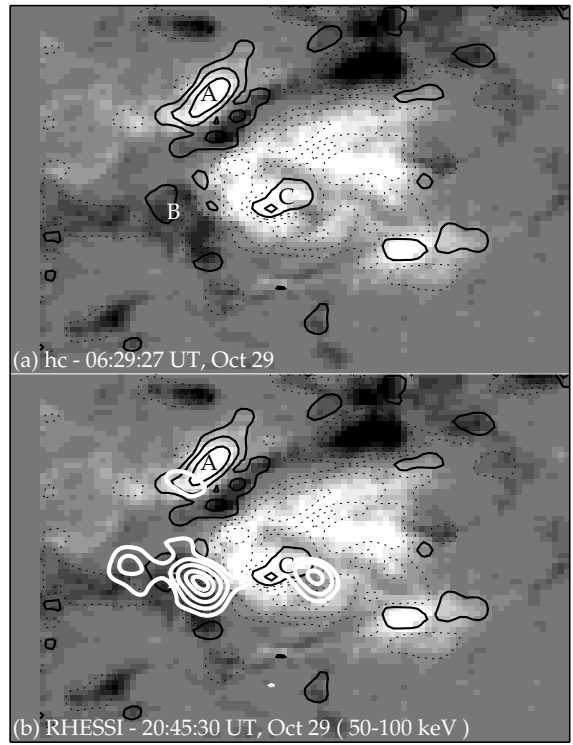}}
\caption{Comparison of the $h_c$ map and the {\sl RHESSI} HXR observations.
({\bf a}) the $h_c$ image at 06:29:27 on Oct. 29, i.e., Figure 3h. ({\bf b})
white contours of the hard X-ray (50-100 keV) emissions during the X10 flare,
superposed on ({\bf a}). The HXR contour levels are 10\%, 20\%, 40\%, 70\%,
90\% of the data maximum. The overlaid RHESSI data is constructed from the time
period 20:45:30 -- 20:46:30 UT. The correlation between the two HXR footpoints
and the counter-helical islands (B and C) is obvious.
\label{fig5}}
\end{figure}

\end{article}
\end{document}